\documentstyle[11pt]{article}

\begin{document}
\begin{center}
{\LARGE{\bf M{\o}ller's Energy in the Kantowski-Sachs Space-time}}\\[2em]
\large{\bf{{ M. Abdel-Megied} and Ragab M. Gad}\footnote{Email Address: ragab2gad@hotmail.com}}\\
\normalsize {Mathematics Department, Faculty of Science,}\\
\normalsize  {Minia University, 61915 El-Minia,  EGYPT.}
\end{center}

\begin{abstract}
We present a counter example to paper \cite{P71}  and show that the
result obtained is correct for a class of metric but not general. We
calculate the  total energy of the Kantowski-Sachs space-time by
using the energy-momentum definitions of M{\o}ller in the theory of
general relativity and the tetrad theory of gravity.
\end{abstract}

\setcounter{equation}{0}
\section{Introduction}
Since the birth of the theory of general relativity and this theory
has been accepted as a superb theory of space-time and gravitation,
as many physical aspects of nature have been experimentally verified
in this theory. However, this theory is still incomplete theory,
namely, it lacks definition of energy and momentum. In this theory
many physicist have introduced different types of energy-momentum
complexes \cite{Many}, each of them being a pseudo-tensor, to solve
this problem. The non-tensorial property of these complexes is
inherent in the way they have been defined and so much so it is
quite difficult to conceive of a proper definition of energy and
momentum of a given system. The recent attempt to solve this problem
is to replace the theory of general relativity by another theory,
concentrated on the gauge theories for the translation group, the so
called teleparallel equivalent of general relativity. We were hoping
that the theory of teleparallel gravity would solve this problem.
Unfortunately, the localization of energy and momentum in this
theory is still an open, unresolved and disputed problem as in the
theory of general relativity.
\par
M{\o{ller modified the theory of general relativity by constructing
a gravitational theory based on Weitzenb\"{o}ck space-time. This
modification was to overcome the problem of the energy-momentum
complex that appears in Riemannian space. In a series of paper
\cite{M1}-\cite{M3}, he was able to obtain a general expression for
a satisfactory energy-momentum complex in the absolute parallelism
space. In this theory the field variable are 16 tetrad components
$h_{a}^{\,\,\,\mu}$, from which the Riemannian metric arises as
\begin{equation}\label{g}
g_{\mu\nu}=\eta_{ab}h^{a}_{\,\,\,\mu}h^{b}_{\,\,\,\nu}.
\end{equation}
\par
The basic purpose of this paper is to obtain the total energy of the
Kantowski-Sachs space-time by using the energy-momentum definitions
of M{\o}ller in the theory of general relativity and the tetrad
theory of gravity.

 The standard representation of
Kantowski and Sachs space-times are given by \cite{KS}

\begin{equation} \label{K-S}
ds^2= dt^2 - A^2(t) dr^2 - B^2(t)\big(d\theta^2 + \sin^2\theta
d\phi^2\big),
\end{equation}
where the functions $A(t)$ and $B(t)$ are function in $t$ and
determined from the field equations.\\
For more detailed descriptions of the geometry and physics of this
space-time see for example \cite{KS}, \cite{KC} and \cite{C77}.

\setcounter{equation}{0}
\section{On the fourth component of Einstein's complex}
Prasanna have shown that space-times with purely time dependent
metric potentials have their components of total energy and momentum
for any finite volume $(T^4_i +t^4_i)$ identically zero. He had used
the Einstein complex for the general Riemannian metric
\begin{equation}\label{RM}
dS^2 -g_{ij}(x^0)dx^idx^j,
\end{equation}
and concluded the  following: For space-times with metric potentials
$g_{ij}$ being functions of time variable alone and independent of
space variable the components $(T^4_i +t^4_i)$
vanish identically as a consequence of conservation law.\\
Unfortunately the conclusion above is not the solution to the
problem considered, in the sense that it does not give the same
result for all metrics have form (\ref{RM}), using Einstein complex.
If (\ref{RM}) is given in spherical coordinates, then Prasanna's
conclusion is correct by using  M{\o}ller's  complex but not correct
for all metrics by using Einstein's complex. Because M{\o}ller's
complex could be utlized to any coordinate system, but Einstein's
complex give meaningful result if it is evaluated in Cartesian
coordinates. In the present paper we have found that the total
energy for the Kantwaski-Sachs space-time is identically zero by
using M{\o}ller's complex, but not zero by using Einstein's complex.

In a recent paper \cite{GF2007}, Gad and Fouad have found the energy
and momentum distribution of Kantowaski-Scahs space-time, using
Einstein, Bergmann-Thomson, Landau-Lifshitz and Papapetrou energy
momentum complexes. In this section we restrict our attention to the
Einstein's  complex which is defined by \cite{E}
\begin{equation}\label{EC}
\theta^{k}_{i} = T^{k}_{i} + t^{k}_{i} = u_{i\,\,\,\,\,\,,k}^{[jk]},
\end{equation}
with
\begin{equation} \label{3.2}
u_{i\,\,\,\,\,\,,k}^{[jk]} = \frac{1}{\kappa} \frac{g_{in}}{\sqrt{-
g}} \big[ - g\big( g^{kn}g^{lm} - g^{ln}g^{km}\big)\big]_{,m}.
\end{equation}
The energy and momentum in the Einstein's prescription are given by
\begin{equation}
P_{i}=\int\int\int \theta^{0}_{i}dx^1 dx^2 dx^3.
\end{equation}
The Einstein energy-momentum complex satisfies the locall
conservation law
\begin{equation} \label{CL}
\frac{\partial\theta^{k}_{i}}{\partial x^k}=0.
\end{equation}
The energy density for the space-time under consideration, in the
Cartesian coordinates, obtained in \cite{GF2007} is
\begin{equation}\label{E-density}
\theta^{0}_{0} =\frac{1}{8\pi Ar^4}(A^2r^2-B^2),
\end{equation}
and the total energy is
$$
E_{Ein}=P_{0}= \frac{1}{2Ar}\big(A^2r^2 + B^2\big),
$$

 Following the approach in \cite{GF2007}, we obtain the
following components of $\theta^{k}_{0}$
\begin{equation}\label{8}
\begin{array}{ccc}
\theta_{0}^{1}& =-\frac{x}{8\pi A^2 r^4}\big[\dot{A}A^2r^2 - B(2A\dot{B}-B\dot{A})\big],\\
\theta_{0}^{2}& =-\frac{y}{8\pi A^2 r^4}\big[\dot{A}A^2r^2 - B(2A\dot{B}-B\dot{A})\big],\\
\theta_{0}^{3}& =-\frac{z}{8\pi A^2 r^4}\big[\dot{A}A^2r^2 -
B(2A\dot{B}-B\dot{A})\big].
\end{array}
\end{equation}
The components (\ref{E-density}) and (\ref{8}) satisfy the
conservation law (\ref{CL}).
\par
Hence from equations (\ref{E-density}) and (\ref{8}), we have
$\theta^{i}_{0} = T^{i}_{0} + t^{i}_{0} \neq 0$ consequently
$\theta^{0}_{0} = T^{0}_{0} + t^{0}_{0}$ is not identically zero.

 \setcounter{equation}{0}
\section{Energy in the theory of General Relativity}
In the general theory of relativity, the energy-momentum complex
of M{\o}ller in a four dimensional background is given as
\cite{M1}
\begin{equation}\label{4.1}
\Im^k_i = \frac{1}{8\pi}\chi ^{kl}_{i,l},
\end{equation}
where the antisymmetric superpotential $\chi^{kl}_i$ is
\begin{equation}\label{4.2}
\chi^{kl}_i = - \chi^{lk}_i = \sqrt{-g}\big( \frac{\partial
g_{in}}{\partial x^m} - \frac{\partial g_{im}}{\partial x^n}\big)
g^{km}g^{nl},
\end{equation}
$\Im^0_0$ is the energy density and $\Im^0_{\alpha}$ are the
momentum density components. \\
Also, the energy-momentum complex $\Im^{k}_{i}$ satisfies the
local conservation laws:
\begin{equation}\label{4.3}
\frac{\partial\Im^k_i}{\partial x^k} = 0
\end{equation}
The energy and momentum components are given by
\begin{equation} \label{4.3-}
P_i=\int\int\int\Im^0_i dx^1dx^2dx^3 =
\frac{1}{8\pi}\int\int\int\frac{\partial\chi^{0l}_{i}}{\partial x^l}
dx^1dx^2dx^3.
\end{equation}
For the line element (\ref{K-S}), the only non-vanishing components
of $\chi^{kl}_{i}$ are
\begin{equation}\label{4.4}
\begin{array}{ccc}
\chi^{01}_{1} & = -\frac{B^2(t)}{A(t)}\sin\theta,\\
\chi^{02}_{2} & = -A(t)\sin\theta,\\
\chi^{03}_{3} & = -\frac{A(t)}{\sin\theta}.
\end{array}
\end{equation}
Using these components  in equation (\ref{4.1}), we get the energy
and momentum densities as following
\begin{equation}
\Im_{0}^{0} = 0.
\end{equation}
\begin{equation}
\Im_{1}^{0} = \Im_{3}^{0}= 0, \Im_{2}^{0}=-A(t)\cos\theta.
\end{equation}
From equation (\ref{4.3-}) and (\ref{4.4}) and applying the Gauss
theorem, we obtain the total energy and momentum components in the
following form
\begin{equation}
P_0=E=0,
\end{equation}
\begin{equation}
P_{\alpha}=0,
\end{equation}

\setcounter{equation}{0}
\section{Energy in the Tetrad Theory of gravity}
 The super-potential of M{\o}ller in the
tetrad theory of gravity is given by (see \cite{M2}- \cite{MW})
\begin{equation} \label{TT}
U_{a}^{bc}=
\frac{\sqrt{-g}}{2\kappa}P^{dbc}_{efh}\big[\Phi^fg^{eh}g_{ad}-\lambda
g_{ad}\gamma^{efh}-(1-2\lambda)g_{ad}\gamma^{hfe}\big],
\end{equation}
where
$$
P^{dbc}_{efh}=\delta^d_e g_{fh}^{bc} + \delta^d_f g_{he}^{bc} -
\delta^d_h g_{ef}^{bc},
$$
with $g_{fh}^{bc}$ being a tensor defined by
$$
g_{fh}^{bc}= \delta_{f}^{b}\delta_{h}^{c} -
\delta_{h}^{b}\delta_{f}^{c},
$$
$\gamma_{abc}$ is the con-torsion tensor given by
\begin{equation}\label{gamma}
\gamma_{abc}=h_{i\mu}h^i_{\,\,\,\nu ;\rho}
\end{equation}
and $\Phi_{a}$ is the basic vector defined by
$$
\Phi_{a}=\gamma^{b}_{\,\,\,ab}.
$$
 The energy in this theory is expressed by the following
surface integral
\begin{equation}\label{ETT}
E=\lim_{r\rightarrow\infty}\int_{r=const.}U_{0}^{0\alpha}n_{\alpha}dS,
\end{equation}
where $n_{\alpha}$ is the unit three vector normal to the surface
element $dS$.

The tetrad components of the space-time (\ref{K-S}), using
(\ref{g}), are as following

\begin{equation}
\begin{array}{ccc}
h^a_{\,\,\, \mu}& = \big[ 1, A(t), B(t), B(t)\sin\theta \big],\\

h_a^{\,\,\, \mu}& = \big[ 1, A^{-1}(t), B^{-1}(t),
\frac{B^{-1}(t)}{\sin\theta} \big].
\end{array}
\end{equation}
Using these components in (\ref{gamma}), we get the non-vanishing
components of $\gamma_{\mu \nu\beta}$ as following
\begin{equation}
\begin{array}{ccc}\label{VG}
\gamma_{011}=-\gamma_{101} &=-A(t)\dot{A}(t),\\
\gamma_{022}=-\gamma_{202} &=-B(t)\dot{B}(t),\\
\gamma_{033}=-\gamma_{303} &=-B(t)\dot{B}(t)\sin^2\theta,\\
\gamma_{233}=-\gamma_{323} &=-B^2(t)\sin\theta \cos\theta .
\end{array}
\end{equation}
Consequently, The only non-vanishing components of basic vector
field are
\begin{equation}\label{phi}
\begin{array}{ccc}
\Phi^0 &=-2\big\{ \frac{\dot{A}(t)}{A(t)} + \frac{\dot{B}(t)}{B(t)}\big\},\\
\Phi^2 &=\frac{\cot\theta}{B^2(t)}.
\end{array}
\end{equation}
Using (\ref{VG}) and (\ref{phi}) in (\ref{TT}) and (\ref{ETT}), we
get
$$
E=0.
$$

\section{Summary and Discussion}

In this paper we have shown that the fourth component of Einstein'
complex for the Kantowski-Sachs space-time is not identically zero.
This give a counter example to the result obtained  by Prasanna
\cite{P71}. We calculated the total energy of Kantowski-Sachs
space-time Using M{\o}ller's tetrad theory of gravity. We found that
the total energy is zero in this space-time. This result do not
agree with the previous results obtained in the both theories of
general relativity \cite{GF2007} and  teleparallel  gravity
\cite{Gijtp}, using Einstein, Bergmann-Thomson and Landau-Lifshitz
energy-momentum complexes. In both theories the energy and momentum
densities for this space-time are finite and reasonable. We notice
that these results are not in conflict with that given by
M{\o}ller's values for the energy and momentum densities if $r$
tends to infinity.



\begin{thebibliography}{99}
\bibitem{P71} A. R. Prasanna, Progress of Theoretical Phsics,
{\bf{45}}, 1330 (1971).
\bibitem{Many}  R. C. Tolman,  "Relativity,
Thermodynamics and Cosmology, (Oxford University Press, Oxford), p.
227 (1934); L. D. Landau  and E. M. Lifshitz, "The Classical Theory
of Fields", (Addison-Wesley Press, Reading, MA) p. 317 (1951);  A.
Papapetrou, Proc. R. Ir. Acad. {\bf{A52}}, 11 (1948);  P. G.
Bergmann  and R. Thompson, Phys. Rev. {\bf{89}}, 400 (1953);  S.
Weinberg, "Gravitation and Cosmology: Principles and Applications of
General Theory of Relativity" ( Wiley, New York) 165 (1972).
\bibitem{M1} C. M{\o}ller,  Ann. Phys. (NY) {\bf{4}}, 347 (1958).
\bibitem{M2} C. M{\o}ller, Mat. Fys. Medd. K. Vidensk. Selsk.
{\bf{31}}, 10 (1961).
\bibitem{M3} C. M{\o}ller, Mat. Fys. Medd. K. Vidensk. Selsk.
{\bf{39}}, 13 (1978).
\bibitem{MW} F. I. Mikhail, M. I. Wanas, A. Hindawi and E. I.
Lashin, Int. J. Theor. Phys., {\bf{32}}, 1627 (1993).
\bibitem{KS} R. Kantowski and R. K. Sachs, J. Math. Phys. {\bf{7}},
443 (1966).
\bibitem{KC} A. S. Kompaneets and Chernov A. S., Sov. Phys. JETP
{\bf{20}}, 1303 (1965).
\bibitem{C77} C. B. Collins, J. Math. Phys. {\bf{18}}, 2116
(1977).
\bibitem{GF2007} R. M. Gad and A. Fouad, Astrophys. Space Sci.,
 {\bf{310}}, 135  (2007).
\bibitem{E} A. Einstein, Sitzungsber. Preuss. Akad. Wiss. Berlin
(Math. Phys.) {\bf{778}} (1915).


\bibitem{Gijtp} R. M. Gad, Int. J. Theor. Phys.  {\bf{46}}, 3263
(2007).


\end{thebibliography}
\end{document}